\documentclass[aps,pra,amsmath,amsfonts,amssymb,showpacs,preprint]{revtex4}

\usepackage{amssymb}
\usepackage{graphicx,psfrag,color}
\usepackage{dcolumn}
\usepackage{bm}
\usepackage{mathbbol}

\begin{document}

\title{Entanglement, identical particles and the uncertainty principle}


\author{Gustavo Rigolin}
\email{rigolin@ufscar.br}
\affiliation{Departamento de F\'isica, Universidade Federal de
S\~ao Carlos, 13565-905, S\~ao Carlos, SP, Brazil}

\date{\today}

\begin{abstract}
A new uncertainty relation (UR) is obtained for a system of N identical pure entangled particles
if we use symmetrized obser\-vables when deriving the inequality. 
This new expression can be written in a form where we identify a term which 
explicitly shows the quantum correlations among the particles that constitute the system. 
For the particular cases of two and three particles, making use of the Schwarz inequality, 
we obtain new lower bounds for the UR that are different from the standard one. 
\end{abstract}

\pacs{03.65.Ta, 03.65.Ud}

\maketitle

\section{Introduction}
One of the most important consequences of Quantum Mechanics (QM) is a theoretical limit imposed 
on the simultaneous measurement of canonically conjugated variables. This limitation, first 
derived by Heisenberg \cite{heisenberg}, departs from the classical belief that we can, 
in principle, reduce the uncertainties in the measurements just by building more and more accurate measurement devices. 

Many experiments have checked, using a variety of canonically conjugated variables, 
the validity of the UR. No one has ever verified a violation of the inequality that 
imposes lower bounds in the product of the dispersions of the canonical variables.

Recently, however, an experiment \cite{kim} has suggested that we could have a 
violation of the UR. A
possible explanation for that ``violation'' is that the Heisenberg uncertainty relation (HUR), 
as we know it, derived for a single particle, cannot be consistently applied when we deal with more than one particle at the same time \cite{rigolin}. 
Indeed,
the data from the experiment of 
Ref. \cite{kim}, and then used to compute the dispersions in position and momentum,
come from simultaneous measurements on two identical and entangled particles. In this sense, there is no violation of the HUR since
in its original derivation the quantum correlation between identical particles were not taken into account.  

In this article we want to formalize this idea by deriving a general UR 
for a pure quantum state describing N identical and entangled particles, which explicitly 
shows the correlations among the particles that constitute the system. Here two particles are considered correlated if 
for some pair of observables the quantum covariance function defined in Eq.~(\ref{QCF1}) is non zero. 
Since we deal with pure states, any possible correlation is called quantum correlation or equivalently entanglement.
Another assumption in the derivation of this general UR is that we are dealing with identical particles. 
Putting together all these pieces, we arrive at a generalized UR that explains the experimental result of Ref. \cite{kim}, 
the main motivation for this paper. 
We also apply the general UR to the case of two and three identical particles, 
where, using the Schwarz inequality, we obtain new lower bounds for the UR valid when we deal with identical and
entangled particles.

\section{Identical and entangled particles}

\subsection{Identical particles}

It is well known that when we deal with a system of N identical particles 
we should use (anti-) symmetrized wave functions to describe an ensemble of (fermions) bosons. 
Less known, however, is the fact that we should also use \textit{physical observables} \cite{cohen,messiah}, defined as those that 
commute with the permutation ope\-rators of the system. Mathematically, a physical observa\-ble must 
satisfy the following commutation relation,
\begin{equation}
[ {\cal O}, P ] = 0,   \label{comutadorzero}
\end{equation}
where ${\cal O}$ is an observable and $P$ is any permutation ope\-rator of the system. 

It is worth mentioning that physical observables are symmetrical operators. 
These operators are the only quantities that should be symmetrical to satisfy 
the symmetrization postulate \cite{cohen,messiah}.  Manipulating them properly we can arrive at any 
physical quantity of interest, being it symmetrical or not. Physical quantities,
for instance, 
are real numbers which are not subjected to the symmetrization postulate. 
Therefore, mean values such as the dispersion in position and momentum for one 
particle, i. e.  $\Delta Q_{1}$ and $\Delta P_{1}$, are legitimate real numbers 
that can be calculated from mathematical manipulations of symmetrical operators.

\subsection{Pure entangled particles}

The Hilbert space $\mathcal{H}$ of a system of N particles is the direct product 
of the Hilbert spaces $\mathcal{H}_{i}$ associated with each particle individually,   
\begin{equation}
\mathcal{H} = \mathcal{H}_{1} \otimes \mathcal{H}_{2} \otimes \ldots \otimes \mathcal{H}_{N} = \bigotimes_{i=1}^{N} \mathcal{H}_{i}. 
\end{equation}

The general state $\left| \Psi \right>$ which describes a pure system is \cite{cohen,isham}
\begin{equation}
\left| \Psi \right>  =  \sum_{i_{1}, \ldots, i_{N}}^{}{c_{i_{1} \ldots i_{N}} \left|u_{i_{1}}\right>_{1} \otimes \ldots \otimes \left|u_{i_{N}}\right>_{N}}, \label{estado}
\end{equation}
where $c_{i_{1} \ldots i_{n}}$ are the expansion coefficients of the state  $\left| \Psi \right>$ 
in the base $\{ \left|u_{i_{1}} \right>_{1},$ $\ldots ,$ $\left|u_{i_{N}}\right>_{N} \}$. 
Here, a ket $\left|u_{i_{n}} \right>_{n}$ means that we have particle $n \leq N$ in the state $\left| u_{i_{n}} \right>$.

A \textit{pure state} describing a system of N particles is considered entangled if we 
cannot write the ket $\left| \Psi \right>$ as a tensor product of N kets, each one belonging to an individual particle. 
It means that we have a non-factorizable state \footnote{Entanglement is a partition dependent concept. 
For example, in the state $|u_{1}\rangle_{1} \otimes (|u_{1}\rangle_{2}|u_{2}\rangle_{3} +|u_{2}\rangle_{2}|u_{1}\rangle_{3})$ 
particles $1$ and $2$ are not entangled while $2$ and $3$ are. 
The definition of entanglement used here considers states like this to be entangled.}
\begin{equation}
\left| \Psi \right> \; \neq \; \left| \psi_{1}\right> \otimes \ldots \otimes \left| \psi_{N}\right>, 
\end{equation}
which implies that Eq.~(\ref{estado}) cannot be written as

\begin{equation}
\left| \Psi \right> \; \neq \; \sum_{i_{1}}^{}{c_{i_{1}} \left|u_{i_{1}}\right>_{1}} 
\otimes \ldots \otimes \sum_{i_{N}}^{}{c_{i_{N}} \left|u_{i_{N}} \right>_{N}},
\end{equation}
where $c_{i_{1}}, \ldots , c_{i_{N}}$ are the expansion coefficients of the non entangled 
state $\left| \psi_{1}\right> \otimes \ldots \otimes \left| \psi_{N}\right>$ 
in the base $\{ \left|u_{i_{1}} \right>_{1}, \ldots, \left|u_{i_{N}}\right>_{N} \}$. 

\section{New uncertainty relation}

\subsection{N particles}

We know that any two operators, $A$ and $B$, satisfy the following inequality,
\begin{equation}
(\Delta A)^{2}(\Delta B)^{2} \geq \frac{|\left< [A, B] \right> |^{2}}{4},  \label{geral}
\end{equation}
where 
\begin{eqnarray*}
\Delta A &=& \sqrt{\langle \psi |A^{2}| \psi \rangle -\langle \psi |A| \psi \rangle^{2}}, \\
\Delta B &=& \sqrt{\langle \psi |B^{2}| \psi \rangle -\langle \psi |B| \psi \rangle^{2}},
\end{eqnarray*} 
are the root mean square deviation (dispersion) of the operators $A$ and $B$, for a given state $|\psi\rangle$, respectively.
 We now require that when deriving the uncertainty relation of an ensemble of 
 N pure identical and entangled particles we should satisfy the conditions stated above about entangled and identical particles,
 namely:
\begin{enumerate}
\item Use only physical observables.
\item Use non-factorizable states.
\end{enumerate}

With only these two assumptions, which represent mathematically,
in the simplest form, the fact that we are dealing with a system of N identical and 
entangled particles, we shall obtain a new uncertainty relation that generalizes the traditional HUR.

Let us begin the derivation of this generalized uncertainty relation (GUR) by defining 
the physical observables that we should use in Eq.~(\ref{geral}).  
In the state space $\mathcal{H}_{i}$ of the particle $i$ we have the usual position and 
momentum observables, $Q_{i}$ and $P_{i}$, respectively. When treating a system of 
N particles we use the extended observables defined below (from now on, we restrict ourselves, 
for simplicity, to the one dimensional case):
\begin{equation}
Q_{i} = {\cal I}_{1} \otimes \ldots \otimes Q_{i} \otimes \ldots \otimes {\cal I}_{N},  \label{extended1}
\end{equation}
\begin{equation}
P_{i} = {\cal I}_{1} \otimes \ldots \otimes P_{i} \otimes \ldots \otimes {\cal I}_{N},  \label{extended2}
\end{equation}
where ${\cal I}_{i}$ is the identity operator in the state space of particle $i$.
It is easily shown that these observables do not commute with the permutation 
operators defined in the state space of the N particle system. We need, then, 
to create a pair of physical observables that should be used in the derivation 
of the uncertainty relation. The simplest pair of physical observables is written as

\begin{equation}
Q = Q_{1} + \ldots + Q_{N},   \label{physical1}
\end{equation} 
\begin{equation}
P = P_{1} + \ldots + P_{N},   \label{physical2}
\end{equation}
where $Q_{i}$ and $P_{i}$ are the extended observables defined in Eqs.~(\ref{extended1}) and (\ref{extended2}).
Using these two physical observables in Eq.~(\ref{geral}) we get
\begin{equation}
(\Delta Q)^{2}(\Delta P)^{2} \geq \frac{| \left< [Q, P] \right> |^{2}}{4}. \label{eq1}
\end{equation} 
But the commutator of $Q$ and $P$ is
\begin{equation}
[ Q, P ] = [Q_{1},P_{1}] + \ldots + [Q_{N},Q_{N}] = iN\hbar.
\end{equation}
Then, Eq.~(\ref{eq1}) becomes
\begin{equation}
(\Delta Q)^{2} (\Delta P)^{2} \; \geq \; \frac{N^{2}\hbar^{2}}{4}. \label{eq2}
\end{equation}

If we define the quantum covariance function (QCF) for the position and for the momentum as \cite{delaTorre}
\begin{equation}
C_{Q}(i,j) =  \left< Q_{i} Q_{j} \right> - \left< Q_{i} \right> \left< Q_{j} \right>, \label{QCF1}
\end{equation}
\begin{equation}
C_{P}(i,j) =  \left< P_{i} P_{j} \right> - \left< P_{i} \right> \left< P_{j} \right>,
\end{equation}
Eq.~(\ref{eq2}) becomes
\begin{equation}
\sum_{i,j=1}^{N}{C_{Q}(i,j)} \sum_{i,j=1}^{N}{C_{P}(i,j)} \; \geq \; 
\frac{N^{2}\hbar^{2}}{4}. \label{geral1}
\end{equation}
But since $C_Q(i,i)= (\Delta Q_{i})^{2}$ and $C_P(i,i)= (\Delta P_{i})^{2}$, we can write Eq.~(\ref{geral1}) as
\begin{equation}
\left( \sum_{i=1}^{N}{(\Delta Q_{i})^{2}} + \sum_{i \neq j=1}^{N}{C_{Q}(i,j)} \right) 
\left( \sum_{i=1}^{N}{(\Delta P_{i})^{2}} + \sum_{i \neq j=1}^{N}{C_{P}(i,j)} \right) \;  \geq  \; \frac{N^{2}\hbar^{2}}{4}. \label{geral2}
\end{equation}
This is the GUR that we should use when working with a system of N entangled and identical 
particles, our main result in this paper. Note that 
since we are dealing with pure states, entanglement manifests itself by the non-vanishing QCF \cite{ballentine}. 
We clearly see that in this expression we get two terms for each physical observable. 
For the physical observable $Q$, $\sum_{i=1}^{N}{(\Delta Q_{i})^{2}}$ is the sum of 
the square of the dispersions in position for each particle that constitutes the system. 
The other one, $\sum_{i \neq j=1}^{N}{C_{Q}(i,j)}$, is the term that shows the quantum 
correlations among all the particles that belong to the system. If we could factorize 
the state describing the system this term would be zero and we would recover the HUR. It is worthy noting that dealing 
with mixed states it is possible to have at the same time a non-vanishing QCF and a separable state. 
Since here we are interested only in pure states, a vanishing QCF always implies separable (non-entangled) states \cite{ballentine}.

Let us show explicitly that we do indeed get the usual HUR if the QCF are all zero. In this case, 
$\sum_{i \neq j=1}^{N}{C_{Q}(i,j)}= 0$ and we have $N$ independent equal particles, namely,
$(\Delta Q_{i})^{2}=(\Delta Q_{j})^{2}$ and $(\Delta P_{i})^{2}=(\Delta P_{j})^{2}$ for any $i$ and $j$. Inserting
these facts into Eq.~(\ref{geral2}) we readily see that it implies $\Delta Q_{i}\Delta P_{i}\geq \hbar/2$, for all particles (any $i$).
This last equation is the usual HUR.


\subsection{Two particles} 

We now restrict the GUR for the case of two particles. In this scenario Eq.~(\ref{geral2}) becomes

\begin{displaymath}
\left[ \frac{(\Delta Q_{1})^{2}}{2}+\frac{(\Delta Q_{2})^{2}}{2}+
(\left< Q_{1}Q_{2} \right> - \left< Q_{1} \right> \left< Q_{2} \right>) \right] \times \;\;\;\;\;\;\;\;\;\;\;
\end{displaymath}
\begin{equation}
\left[ \frac{(\Delta P_{1})^{2}}{2}+\frac{(\Delta P_{2})^{2}}{2}+
(\left< P_{1}P_{2} \right> - \left< P_{1} \right> \left< P_{2} \right>) \right]
 \geq \frac{{\hbar}^{2}}{4}. \label{casotwo1}
\end{equation}

We simplify further Eq.~(\ref{casotwo1}) invoking the Schwarz inequality,
\begin{equation}
\left< \varphi_{1} \mid \varphi_{1} \right> \left< \varphi_{2} \mid \varphi_{2}
\right> \; \geq \; \left< \varphi_{1} \mid \varphi_{2} \right> \left< \varphi_{2} \mid \varphi_{1} \right>, \label{schwarz}
\end{equation} 
where $\left| \varphi_{1} \right>$ and $\left| \varphi_{2} \right>$ are two generic states. 
Since Eq.~(\ref{schwarz}) is valid for any state, it is valid for the two states below,
\begin{equation}
\left| \varphi_{1} \right> \; = \; (Q_{1} \pm Q_{2}) \left| \psi \right>,
\end{equation}
\begin{equation}
\left| \varphi_{2} \right> \; = \; \left| \psi \right>,
\end{equation} 
where $\left| \psi \right>$ represents the normalized wave function that describes our system of two particles. Hence, applying Eq.~(\ref{schwarz})
for these two states we get
\begin{equation}
\mid \left< Q_{1}Q_{2} \right> - \left< Q_{1} \right> \left< Q_{2} \right> \mid \; \leq \; \frac{(\Delta Q_{1})^{2}}{2} + \frac{(\Delta Q_{2})^{2}}{2}. \label{essa1}
\end{equation}  
The same reasoning can be used for the momentum ope\-rator and we get a similar expression,
\begin{equation}
\mid \left< P_{1}P_{2} \right> - \left< P_{1} \right> \left< P_{2} \right> \mid \; \leq \; \frac{(\Delta P_{1})^{2}}{2} + \frac{(\Delta P_{2})^{2}}{2}. \label{essa2}
\end{equation}
Analyzing Eqs.~(\ref{essa1}) and (\ref{essa2}) we see that the absolute value of the QCF 
for the position and for the momentum is always less than the mean of the dispersions in 
position and momentum of the two particles. Hence, we have an upper bound for the QCF. 
Note that if we are dealing with non-entangled states the left hand side of Eqs.~(\ref{essa1}) 
and (\ref{essa2}) are always zero. In this situation Eq.~(\ref{casotwo1}) is equivalent to 
the usual HUR and furnishes no new result. In other words, entanglement is essential to 
obtain non trivial upper bounds for the QCF which imply the new results that follow. 

Substituting these upper bounds in Eq.~(\ref{casotwo1}) we get
\begin{equation}
\left[ (\Delta Q_{1})^{2}+(\Delta Q_{2})^{2} \right] 
\left[ (\Delta P_{1})^{2}+(\Delta P_{2})^{2} \right]
\geq \frac{{\hbar}^{2}}{4}. \label{casotwo2}
\end{equation} 
This last expression should be the correct uncertainty relation when treating an entangled pair of identical particles. 

An interesting case arises when we produce an entangled pair of identical particles with 
the same dispersion in position and in momentum. Using these 
assumptions, 
i. e. $\Delta Q_{1}=\Delta Q_{2}$ and $\Delta P_{1}=\Delta P_{2}$, Eq.~(\ref{casotwo2}) becomes
\begin{equation}
\Delta Q_{i} \Delta P_{i}  \geq  \frac{{\hbar}}{4}, \label{uqa2}
\end{equation} 
where i=1,2. Here we see that the GUR can be twice smaller than the traditional HUR.



\subsection{Three particles}

For three particles Eq.~(\ref{geral2}) reads
\begin{equation}
\left( \sum_{i=1}^{3}{(\Delta Q_{i})^{2}} + \sum_{i \neq j=1}^{3}{C_{Q}(i,j)} \right) \times
\left( \sum_{i=1}^{3}{(\Delta P_{i})^{2}} + \sum_{i \neq j=1}^{3}{C_{P}(i,j)} \right) 
\; \geq \; \frac{9\hbar^{2}}{4}. \label{casothree1}
\end{equation}
Again we can apply the Schwarz inequality to simplify the above expression. We now 
substitute the following two states in Eq.~(\ref{schwarz}),
\begin{equation}
\left| \varphi_{1} \right> = (a_{1}Q_{1} + a_{2}Q_{2} + a_{3}Q_{3}) \left| \psi \right>,
\end{equation}
\begin{equation}
\left| \varphi_{2} \right> = \left| \psi \right>,
\end{equation}
where $a_{i}= \pm 1, i=1,2,3$. \\
Thus, the Schwarz inequality becomes
\begin{equation}
\left< (a_{1}Q_{1} + a_{2}Q_{2} + a_{3}Q_{3})^2 \right>  \geq  
\left< (a_{1}Q_{1} + a_{2}Q_{2} + a_{3}Q_{3}) \right>^{2}. \label{s1}
\end{equation}
Manipulating Eq.~(\ref{s1}) and an equivalent expression for the momentum 
we get the following inequality (see Appendix \ref{apA}):
\begin{equation}
\left( \sum_{i=1}^{3}{(\Delta Q_{i})^{2}} \right) 
\left( \sum_{i=1}^{3}{(\Delta P_{i})^{2}} \right) \; \geq \;
\frac{9\hbar^{2}}{64}.  \label{casothree2}
\end{equation}
This is the GUR that we should use when working with three identical entangled particles.
An interesting case arises when all the three identical and entangled particles are prepared 
with the same dispersions in position and in momentum. Therefore, if
\begin{equation}
\Delta Q_{1} = \Delta Q_{2} = \Delta Q_{3}, \label{Qzao}
\end{equation}
\begin{equation}
\Delta P_{1} = \Delta P_{2} = \Delta P_{3}, \label{Pzao}
\end{equation}
Eq.~(\ref{casothree2}) becomes
\begin{equation}
\Delta Q_{i} \Delta P_{i} \; \geq \; \frac{\hbar}{8}, \label{uqa3} 
\end{equation}
where i=1,2,3. \\
In this case the departure from HUR is more drastic. We have a lower bound in the 
product of the dispersions four times smaller than that furnished by the HUR. 
Here, the conditions given in Eqs~(\ref{Qzao}) and (\ref{Pzao}) are also achieved 
via a triple coincidence measurement on a maximally entangled tripartite state. 

\section{The Kim and Shih's experiment \cite{kim}}
\label{experimentKS}

The experiment of Ref. \cite{kim} 
dealt with a quantum system satisfying the two assumptions we used to arrive 
at the GUR. Indeed, in Ref. \cite{kim} 
identical and entangled pairs of photons were produced by spontaneous parametric down conversion (SPDC)
with the subsequent measurement of their position and momentum in such a way that the quantum correlation between them
was not destroyed, which was achieved by the use of the ``ghost image'' experimental technique \cite{shih}.  
Also, coincidence circuits were employed to guarantee that
the detection of two photons corresponded to the two entangled photons produced in the SPDC process and 
only events (two photon detection) within a very narrow window of time (coincidence) were selected as valid events. 

Note that without the coincidence measurements, we would have spurious events, those in which we are detecting pairs of photons that are not correlated.
In this case, i.e., without post-selecting the valid events, no correlation exists between the photons and we would get the usual 
HUR. Also, if we let both particles/photons interact with something else in an uncontrollable way 
we destroy the entanglement/correlation between them (or average the correlation out) and we should also recover the HUR. 

Our goal now is to analyze the experiment of Ref. \cite{kim} in the light of the new GUR here presented. 
Looking at Fig. 5 of Ref. \cite{kim}, we note two sets of data/curves. The wider curve is the standard single-slit diffraction 
pattern for a single particle, which can be employed to infer the particle's dispersion/uncertainty in momentum when it is localized to
a precision given by the width of the slit (the dispersion/uncertainty in position). 
This curve was obtained sending pairs of photons, created as described above, to physical slits of width $0.16$ mm.
A detector $D_1$, after the slit receiving photon 1, was fixed at the position zero (center of the slit)
while detector $D_2$, after the slit receiving photon 2, was scanned along a direction perpendicular to the momentum of the incoming photon.
The data was obtained only when coincidence events occurred (simultaneous detection of photons at $D_1$ and $D_2$) and
they can be used to infer the dispersion/uncertainty in the momentum perpendicular to the direction of incidence of photon 2.
Note that in this case \textit{both photons interact in an uncontrollable way} with their slits, destroying any possible quantum correlation they might have had
prior to entering the slits. 
That is why the usual HUR applies and, according to the authors of Ref. \cite{kim}, 
their data is compatible with the minimum value of the HUR,
i.e., $\Delta Q_2 \Delta P_2 = \hbar/2$.

The other curve, the narrower one, was obtained in such a way that all the conditions listed here to the validity of the GUR were satisfied. 
Indeed, we have identical and entangled particles and, most importantly, a situation where the quantum correlation between those photons are not
destroyed. This is achieved by sending only one of the photons, photon 1, to a physical slit. Photon number 2, on the other hand, is sent to
a virtual slit, what the authors of Ref. \cite{kim} called the ``ghost image'' of the physical slit of photon 1. 
This virtual slit is achieved by properly inserting a lens in the optical path of the photons in such a way that it creates an image of the physical slit of photon 1 exactly in 
the position where the physical slit 
of photon 2 was located when the data of the wider curve was obtained. According to the authors of 
Ref. \cite{kim},  in this arrangement both photons uncertainty in position are still given by the width of the slit,
even though photon 2 does not pass through a physical slit. We can see that this is true because the photons emerging the SPDC process 
are entangled in momentum \cite{kim}.  Moreover,
since photon 2 does not interact with a physical slit, it is still correlated to photon 1 and in this case we should 
apply the GUR and not the HUR to analyze the product of the dispersions/uncertainties in position and momentum of photon 2. In other
words, the GUR predicts that $\Delta \tilde{Q}_2 \Delta \tilde{P}_2 \geq \hbar/4$. We have used the tildes over $Q_2$ and $P_2$ in order to 
differentiate the dispersions associated to the narrower curve from the wider one.  

Returning to the data given by Fig. 5 of Ref. \cite{kim}, we can estimate $\Delta \tilde{Q}_2 \Delta \tilde{P}_2$ for the actual
experiment if we note that the uncertainty in momentum is proportional to the spread of the diffraction pattern \cite{fey}.
Following \cite{fey}, we set the uncertainty in momentum as proportional to the position of the first minimum of the sinc-square functions
fitting the data. For the wider curve, the first minimum is located about $2.15$ mm while for the narrower one it is roughly at $1.25$ mm.
Therefore, $\Delta \tilde{P}_2/\Delta P_2 \approx 1.25/2.15$. Now, noting that for the wider curve we have $\Delta Q_2 \Delta P_2 = \hbar/2$
and that $\Delta Q_2 = \Delta \tilde{Q}_2$, we immediately get $\Delta \tilde{Q}_2 \Delta \tilde{P}_2 \approx 0.29 \hbar$. This number 
is slightly
higher than $\hbar/4=0.25\hbar$, the minimum value predicted by the GUR, and clearly lower than the minimum value allowed by the HUR, i.e., $0.5 \hbar$. 

Finally, it is worth mentioning that the GUR presented here also predicts that for the first experiment all QCF's should be zero,
i.e., $\left< Q_{1}Q_{2} \right> - \left< Q_{1} \right> \left< Q_{2} \right>=0$ and $\left< P_{1}P_{2} \right> - \left< P_{1} \right> \left< P_{2} \right>=0$.
For the second experiment, however, at least the QCF for the momenta must be non-zero, since we have entanglement between the momenta of the photons.
Also, the values of all QCF's must be compatible with Eq. (\ref{casotwo1}). 
Unfortunately, with the data provided in Ref. \cite{kim} we cannot compute those functions.

\section{Discussion}

The above results deserve careful attention. They cannot be considered a violation 
of the Heisenberg uncertainty principle. We still cannot measure simultaneously 
the position and momentum (or any pair of canonically conjugated observables) of 
a particle. What we have actually shown was that if we prepare a system of identical and entangled 
particles and use coincidence circuits to detect all the constituents of the system in a way that do not 
destroy the correlation between the particles, we can obtain $\Delta Q_{1} \Delta P_{1}  <  \frac{\hbar}{2}$ for the product of 
the dispersions in position and in momentum for at least one of the particles. We should keep in 
mind that the GUR, which implies  for some special cases $\Delta Q_{1} \Delta P_{1}  <  \frac{\hbar}{2}$, 
was deduced in the framework of QM, with no additional postulates. 

In other words, QM does not forbid  $\Delta Q_{1} \Delta P_{1}  <  \frac{\hbar}{2}$ 
for the cases of identical and entangled particles in coincidence measurements.  
All the well known previous deductions of UR's were made considering only an isolated particle. 
What we have reinforced here is that things are not so simple and straightforward when we have 
two or more particles that are quantum correlated. Our results show that what is believed to be valid for 
one particle cannot be always trivially extended to systems of more than one particle in some special experimental conditions.

It is also worth mentioning that possibly there exists one experimental confirmation of the 
GUR by Kim and Shih \cite{kim}, which motivated this article. We wanted to find a theoretical 
explanation to the results of Ref. \cite{kim} 
without invoking any particularity of the experimental setup 
but the fact that we have a pure bipartite entangled system of two identical photons and 
coincidence measurements that do not destroy or average out the quantum correlation between the entangled photons.

Finally, we insist that new experiments with more than two particles or with other kinds of canonically conjugated operators are most welcome. 

\section{Conclusion}

In this article we tried to show that when dealing with N identical and entangled particles we should use a new UR. 
This new UR is a natural consequence of two reasonable assumptions we make when treating such system, that is: 
use only non-factorizable states and physical observables to deduce the UR. 

The first assumption is a definition of entanglement for pure states \cite{cohen,isham} and the second one 
appears because we are dealing with identical particles \cite{cohen,messiah}.

In some special conditions we have shown that for two and three particles it is possible to obtain UR's that 
have different lower bounds than that furnished by the traditional HUR, reinforcing that in QM the physics of 
systems of more than one particle is not a trivial extension of the physics of single particle systems.

\section*{Acknowledgments}
The author acknowledges important suggestions, encouragement and useful discussions from C. O. Escobar and L. F. Santos. 
The author also thanks the anonymous referee for interesting suggestions that led to Sec. \ref{experimentKS}.
This research was supported by Funda\c{c}\~ao de Amparo \`{a} Pesquisa do Estado de S\~ao Paulo (FAPESP).

\appendix

\section{Proof of Eq. (\ref{casothree2})}
\label{apA}

We now present a detailed deduction of Eq.~(\ref{casothree2}). 
First, we expand  Eq.~(\ref{s1}) using the fact that $a_{i}^{2} = 1$. This leads to
\begin{equation}
\sum_{i=1}^{3}{\left< Q_{i}^{2} \right>} + \sum_{i \neq j=1}^{3}{a_{i}a_{j} 
\left< Q_{i}Q_{j} \right>}  \geq  
\sum_{i=1}^{3}{\left< Q_{i} \right>^{2}} + \sum_{i \neq j =1}^{3}{a_{i}a_{j} 
\left< Q_{i} \right> \left< Q_{j} \right>}. \label{s2} 
\end{equation}
Using the fact that  $(\Delta Q_{i})^{2} = \left< Q_{i}^{2} \right> - 
\left< Q_{i} \right>^{2}$, Eq.~(\ref{s2}) reduces to
\begin{equation}
\sum_{i=1}^{3}{(\Delta Q_{i})^{2}} \; \geq \; - \sum_{i \neq j=1}^{3}{a_{i}a_{j}C_{Q}(i,j)}. \label{s3}
\end{equation}
For $a_{1}=1$ and $a_{2}=a_{3}=-1$  Eq. (\ref{s3}) is
\begin{equation}
\sum_{i=1}^{3}{(\Delta Q_{i})^{2}} \; \geq \; 2 \left( C_{Q}(1,2) + C_{Q}(1,3) 
- C_{Q}(2,3) \right). \label{s4}
\end{equation}
For $a_{2}=1$ and $a_{1}=a_{3}=-1$  Eq. (\ref{s3}) is
\begin{equation}
\sum_{i=1}^{3}{(\Delta Q_{i})^{2}} \; \geq \; 2 \left( C_{Q}(1,2) - C_{Q}(1,3) 
+ C_{Q}(2,3) \right). \label{s5}
\end{equation}
For $a_{3}=1$ and $a_{1}=a_{2}=-1$  Eq. (\ref{s3}) is
\begin{equation}
\sum_{i=1}^{3}{(\Delta Q_{i})^{2}} \; \geq \; 2 \left( - C_{Q}(1,2) + C_{Q}(1,3) + C_{Q}(2,3) \right). \label{s6}
\end{equation} 
Adding Eqs. (\ref{s4}), (\ref{s5}), and ({\ref{s6}) we get
\begin{equation}
2 \left( C_{Q}(1,2) + C_{Q}(1,3) + C_{Q}(2,3) \right) \; \leq \; 
3 \sum_{i=1}^{3}{(\Delta Q_{i})^{2}}.  \label{s7}
\end{equation}
Now, for $a_{1}=a_{2}=a_{3}=1$ Eq.~(\ref{s3}) is
\begin{equation}
2 \left( C_{Q}(1,2) + C_{Q}(1,3) + C_{Q}(2,3) \right) \; \geq \; 
- \sum_{i=1}^{3}{(\Delta Q_{i})^{2}}. \label{s8} 
\end{equation}
Combining Eqs. (\ref{s7}) and ({\ref{s8}) we get the following expression
\begin{equation}
- \sum_{i=1}^{3}{(\Delta Q_{i})^{2}} \; \leq \; 
\sum_{i \neq j=1}^{3}{C_{Q}(i,j)} \; \leq \;  
3 \sum_{i=1}^{3}{(\Delta Q_{i})^{2}}.  \label{s9}
\end{equation}

A similar procedure leads to an equivalent expression for the momentum:
\begin{equation}
- \sum_{i=1}^{3}{(\Delta P_{i})^{2}} \; \leq \; 
\sum_{i \neq j=1}^{3}{C_{P}(i,j)} \; \leq \; 
3 \sum_{i=1}^{3}{(\Delta P_{i})^{2}}.  \label{s10}
\end{equation}
These two inequalities furnish upper bounds for the QCF's that appear in the GUR for three identical particles. 
Substituting the upper bounds of Eqs.~(\ref{s9}) and (\ref{s10}) in Eq.~(\ref{casothree1}) we arrive at Eq.~(\ref{casothree2}),
\begin{equation}
\left( \sum_{i=1}^{3}{(\Delta Q_{i})^{2}} \right) 
\left( \sum_{i=1}^{3}{(\Delta P_{i})^{2}} \right) \; \geq \;
\frac{9\hbar^{2}}{64}. 
\end{equation}

\end{document}